\documentclass[pre,showpacs,floatfix,twocolumn]{revtex4-1}

\usepackage{graphicx}
\usepackage{bm}

\usepackage[applemac]{inputenc}

\usepackage{amsmath}
\usepackage{amssymb}
\usepackage{latexsym}
\usepackage{multirow}

\usepackage{vmargin} 
\setmarginsrb{2.5cm}{1cm}{3cm}{1cm}{1cm}{2cm}{1cm}{2cm}  
\usepackage{color}
\definecolor{dockerblue}{rgb}{0.11,0.56,0.98}
\definecolor{oneblue}{rgb}{0,0,0.75}

\begin{document}

\title{Universal time dependent dispersion properties for diffusion in a one-dimensional critically tilted potential}

\author{T. Gu\'erin}
\author{D. S. Dean}
\affiliation{Laboratoire Ondes et
Mati\`ere d'Aquitaine (LOMA), CNRS, UMR 5798 / Universit\'e de  Bordeaux, F-33400 Talence, France}

\bibliographystyle{apsrev}

\begin{abstract}
We consider the time dependent dispersion properties of overdamped tracer particles diffusing in a one dimensional periodic potential under the influence of an additional constant tilting force $F$. The system is studied in the region where the force is close to the critical value $F_c$ at which the barriers separating neighboring potential wells disappear. We show that, when $F$ crosses the critical value, the shape of the Mean-Square Displacement (MSD) curves is strongly modified. We identify a diffusive regime at intermediate time scales, with an effective  diffusion coefficient which is much larger than the late time diffusion coefficient for $F>F_c$, whereas for $F<F_c$ the late time and intermediate time diffusive regimes are indistinguishable. Explicit asymptotic regimes for the MSD curves are identified at all time scales. 
\end{abstract}

\maketitle

In a variety of physical systems, the motion of tracer particles can be described by Fokker-Planck equations or their associated stochastic differential, or Langevin,  equations  \cite{VanKampen1992,oksendal2003stochastic,gardiner1983handbook}. In such systems, the motion results from the combined action of deterministic, the so called drift, and stochastic forces whose amplitudes are given by a diffusivity, or local diffusion constant. 
In a periodic heterogeneous medium, in which the local transport coefficients are constant in time, but spatially periodic, the late time large scale dynamics can be characterized by a mean velocity and an effective diffusion tensor which characterize the mean drift and the spatial extension 
of a cloud of initially close tracer particles. These effective transport coefficients can be widely different from  typical microscopic ones, and 
are important in a number of phenomena such as mixing, pollutant spreading or chemical reactions \cite{leBorgne2013stretching,dentz2011mixing,barros2012flow,brusseau1994transport,Condamin2007}.

The late time effective transport coefficients have been theoretically characterized in a number of systems. For example, they can be calculated in the case of tracer particles diffusing in incompressible hydrodynamic flows \cite{taylor1953dispersion,Shraiman1987,rosenbluth1987effective,McCarty1988,majda1999simplified,dean2001} or in 
porous media \cite{brenner1980dispersion,rubinstein1986dispersion,quintard1994convection,alshare2010modeling,souto1997dispersion,quintard1993transport}
using homogenization theory. In another context, dispersion properties were derived  using methods of statistical physics for particles diffusing in periodic potentials with uniform molecular diffusivity \cite{dean1994,dean2007effective,derrida1983velocity,zwanzig1988diffusion,deGennes1975brownian,lifson1962self}. 
The problem of dispersion in one-dimensional (1D) systems was also investigated at length ~\cite{reimann2001giant,lindner2001optimal,reimann2002diffusion,reimann2008weak,lindner2002,reguera2006entropic,burada2008entropic,lindner2016giant,costantini1999threshold,lindenberg2007dispersionless,lindenberg2005transport,sancho2010rich,marchenko2014particle}. A remarkable prediction in the case of diffusion in a one-dimensional tilted potential is a huge increase of the effective diffusivity when the force approaches a critical value  
\cite{reimann2001giant}, a phenomenon observed in various experiments  \cite{Evstigneev2008,lee2006giant,tierno2010giant,ma2015colloidalSoftMatt} which was recently used to estimate the energy barrier  opposing the steps of a rotary molecular motor \cite{hayashi2015giant}. More recently, analytical results demonstrated how an external force can influence dispersion in periodic systems in higher dimensions \cite{guerin2015}.

These results, however, characterize only the late-time effective diffusivity, whereas dispersion can also be characterized by the time-dependent Mean-Square-Displacement (MSD) which is routinely measured for example in single particle tracking experiments.
The general theory developed in Ref.~\cite{guerin2015} shows that, starting from steady state 
initial conditions, the average drift is independent of the time whereas the MSD actually
evolves in time, eventually attaining the late time diffusive limit.
The approach to the diffusive limit has been calculated in equilibrium systems  \cite{Dean2014PRE,dean2014approach}. Apart from approximate forms of the distribution of particles \cite{SalgadoGarcia2007PRE,kulikov2011gaussian} in tilted potentials, little is known about the temporal evolution of  dispersion coefficients in general non-equilibrium periodic media. 


Recently, a very general formalism was proposed to calculate the MSD in a wide class of non-equilibrium systems \cite{guerin2015,guerin2015kubo}. This formalism has not yet been used to examine the full temporal behavior of dispersion, here we use it to calculate analytically the time-dependent MSD of overdamped particles diffusing in a 1D tilted periodic potential. We derive exact and explicit asymptotic expressions for the dispersion at various time scales in the case where the external force is close to its critical value (at which the barriers between successive potential wells disappear). In earlier studies \cite{reimann2002diffusion,reimann2001giant}, it has been shown that, when the tilting force is close to its critical value, the tracer particles spend most of their time in a very narrow window of positions, and thus the effective diffusivity takes a universal form depending only on the properties of the potential near these positions. Here we will see that this property carries over  to the time-dependent MSD, which admits simple and universal forms at various time scales. 

The outline of the paper is as follows. The model is briefly introduced in Section \ref{SectionModel}. In Section \ref{SectionGeneralFormula}, we derive a formula for the MSD in terms of first passage time densities. This formula is analyzed asymptotically in Section \ref{SectionAsymptotics}, where we show that the shape of the MSD curves show a remarkable change when $F$ crosses its critical value. Our predictions are in excellent agreement with results obtained by simulating the stochastic trajectories.

\section{Model and quantities of interest}
\label{SectionModel}

\begin{figure}[h!]
\includegraphics[width=7cm]{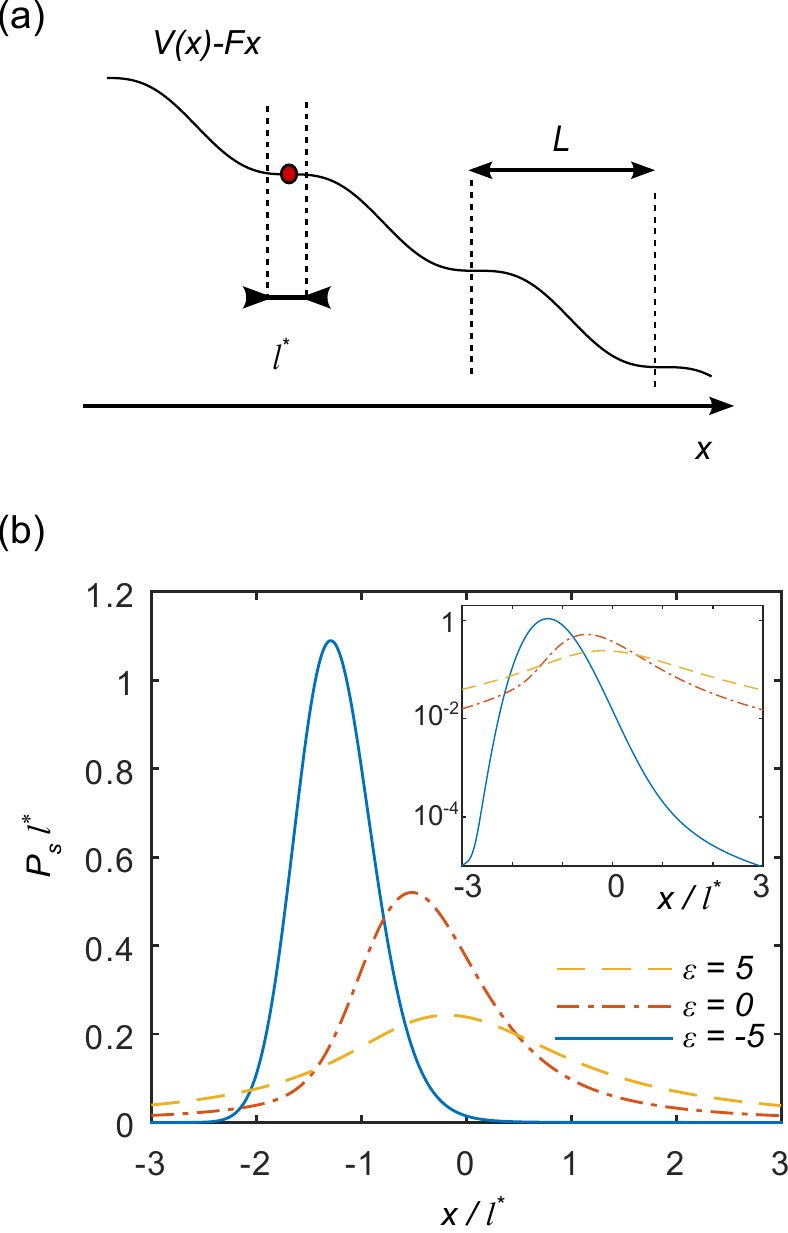} 
 \caption{(a) Sketch of the problem investigated in this paper: we calculate the dispersion properties of particles diffusing in periodic tilted potentials. When the noise is weak, and the force is close to its critical value, the tracer particles spend most of their time in  narrow regions, termed the slow regions, of typical size $l^*$ much smaller than the period $L$. (b) Value of the stationary PDF $P_s(x)$ inside the slow region  obtained by evaluating Eqs.~(\ref{ExpressionPstat}) and (\ref{FluxRescaled}), for $\varepsilon=-5$ ($F$ below the critical force $F_c$), $\varepsilon=0$ ($F=F_c$) and $\varepsilon=5$ ($F>F_c$).  Inset: same curves in semi-logarithmic scale.  
\label{Fig1}}
\end{figure}

We consider the motion of an overdamped tracer particle of position $X(t)$ at time $t$  in a 1D space, moving in a periodic potential $V(x)$ (of period $L$, with $x$ the spatial coordinate) and subject to an additional external  {\em tilting} force $F$ at finite temperature $T$  [see Fig.\ref{Fig1}(a)]. The over-damped dynamics satisfies the force balance equation, which in Langevin form reads
\begin{align}
	\zeta\partial_t X(t)=-V'(X(t))+F+\xi(t),\label{EqLangevin}
\end{align}
where $\zeta$ is the frictional drag coefficient, and the thermal fluctuating forces $\xi(t)$ have zero mean white noise Gaussian statistics characterized by the correlation function $\langle\xi(t)\xi(t')\rangle = 2 k_BT \zeta\delta(t-t')$. We denote $D_0=k_BT/\zeta$ the local molecular diffusivity of the tracer particle, and we define  the drift field
\begin{align}
	u(x)=\zeta^{-1}[-\partial_x V(x)+F].
\end{align}
Equivalently, the process can be described by the Fokker-Planck equation 
\begin{align}
\partial_t P= - \partial_x[u(x)P]+D_0\partial_x^2P,\label{FKPEq}
\end{align}
where $P(x,t)$ is the probability density function of particles at positions $x$ at time $t$.

In this paper, we aim to calculate the time-dependent dispersion quantified by the MSD function $\psi(t)$ defined as
\begin{align}
\psi(t)=\langle [X(t)-X(0)]^2\rangle -\langle X(t)-X(0)\rangle^2,\label{DefMSD}
\end{align}
where $\langle\cdot\cdot\cdot\rangle$ denotes ensemble averaging over realizations of the
white noise. Note that we assume here that the system has reached a steady state at time $t=0$ (in the sense that the probability distribution over a unit cell of one period is the stationary one). 

\section{General expression of the Mean-Square Displacement}
\label{SectionGeneralFormula}
Our starting point is the following Kubo formula, derived in Refs.~\cite{guerin2015,guerin2015kubo}: 
\begin{align}
\hat{\psi}(s)=\frac{2D_0}{s^2}
-\frac{2}{s^2}\int_0^{L} dx\int_0^{L}dx_0 \  u(x) u^*(x_0)\nonumber\\   \left[\hat{P}(x,s\vert x_0)-\frac{P_{s}(x)}{s}\right]P_{s}(x_0). \label{StartingEqMSD}
\end{align}
Here, $\hat{\psi}(s)$ is the Laplace transform of the MSD, \textit{i.e.} $\hat{\psi}(s)=\int_0^{\infty} \psi(t)e^{-st}dt$, while $P(x,t\vert x_0)$ is the propagator of the process modulo $L$ (with periodic boundary conditions), \textit{i.e.} the probability density function for the tracer particle at $x$ (modulo $L$) at $t$ given an initial position $x_0$ (also modulo $L$), and $\hat{P}$ is its temporal Laplace transform. Moreover, $P_{s}(x)=\lim_{t\rightarrow\infty}P(x,t\vert x_0)$ is the probability density function of the position (modulo $L$) in the steady state. 
Note that $P_s$ is not the equilibrium-Boltzmann distribution which is only applicable for a finite  systems with reflecting, or confining, rather than periodic boundary conditions. Here it describes a non-equilibrium steady state , and is characterized by a non-zero flux  $J_{s}$, which has  both a convective and a diffusive (Fickian) component: 
\begin{align}
	J_{s}=u(x) P_{s}(x)-D_0 \partial_x P_{s}(x). \label{DefJs}
\end{align}
Note that $J_{s}$ must be constant in a 1D problem. Finally, in Eq.~(\ref{StartingEqMSD}), $u^*(x)$ represents the drift of the time-reversed stochastic process  \cite{guerin2015kubo},
\begin{align}
	u^*(x)=u(x)- 2J_{s}/P_{s}(x).
\end{align} 

The first step of the present analysis consists in expressing the MSD in terms of First-Passage Times (FPT) densities rather than propagators, this will greatly simplify the asymptotic analysis in the next sections. Consider $f(x,t\vert x_0)$ the  probability density of reaching the position $x$ (modulo $L$) for the first time at $t$, starting from the initial position $x_0$.  The propagators and the FPT densities can be linked by the following, well known,  {\em renewal} equation \cite{VanKampen1992}:
\begin{align}
P(x,t\vert x_0)=\int_0^t dt' f(x,t'\vert x_0)P(x,t-t'\vert x). \label{Renewal}
\end{align}
Physically, the above equation states that if a particle reaches $x$ at $t$, it also means that $x$ was reached for the first time at some earlier instant $t'$, and that  the particle subsequently reached $x$ again in a time $t-t'$. The Laplace transform of Eq.~(\ref{Renewal}) is:
\begin{align}
	\hat{P}(x,s\vert x_0)=\hat{f}(x,s\vert x_0)\hat{P}(x,s\vert x). \label{90243}
\end{align}
We also consider the FPT density averaged over initial conditions,
\begin{align}
	f_{s}(x,t)\equiv\int_0^{L} dx_0 \ f(x,t\vert x_0)P_{s}(x_0),
\end{align}
and we remark that averaging (\ref{90243}) over the stationary distribution for $x_0$ gives
\begin{align}
	\frac{P_{s}(x)}{s}=\hat{f}_{s}(x,s)\hat{P}(x,s\vert x).
\end{align}
Using the above expressions, the equation (\ref{StartingEqMSD}) for the MSD becomes
\begin{align}
\hat{\psi}(s)=\frac{2D_0}{s^2}
- \frac{2}{s^3}\int_0^{L}dx\int_0^{L}dx_0  u(x) u^*(x_0)\nonumber\\
P_{s}(x_0) P_{s}(x)\left[\frac{\hat{f}(x,s\vert x_0)}{\hat{f}_{s}(x,s)}-1\right]. \label{StartingMSD}
\end{align}
This expression for the MSD is well adapted for the asymptotic analysis of dispersion presented in the next section.

\section{Dispersion in a critical tilted potential at different time scales}
\label{SectionAsymptotics}
\subsection{Regions of fast and slow motion}
We now focus on the case where the external force is very close to the critical tilt force 	$F_c=\max[V'(x)]$ [Fig.\ref{Fig1}(a)]. When the force $F\simeq F_c$, the late time effective diffusivity varies as $D_0^{1/3}$ \cite{reimann2002diffusion,reimann2001giant}, and can therefore be much larger than the molecular diffusivity $D_0$ when the latter is small. Our goal here is to predict the approach to this diffusive limit, which is universal  in the sense that it does not depend on the detailed shape of the potential. Following the notations of Refs.~\cite{reimann2002diffusion,reimann2001giant}, we introduce the parameter $\mu$ defined as
\begin{align}
	\mu=-V'''(0)/6,
\end{align}
where we chose the origin of coordinates such that $V'(x)$ is maximum at $x=0$. Note that we assume here that the potential admits a third derivative, more general cases could be treated with our approach but are not considered here for simplicity. We also introduce a parameter $\varepsilon$ to measure the distance to the critical force,
\begin{align}
	\varepsilon= \frac{F-F_c}{\mu^{1/3}(k_BT)^{2/3}}.
\end{align}
The limit considered here is that of weak noise, $k_BT\rightarrow0$ while keeping the parameter $\varepsilon$ constant. In this limit, since $D_0\propto k_BT$, the convective terms dominate over the diffusive terms in  Eq.~(\ref{FKPEq}) almost everywhere except in a narrow region of characteristic size $l^*$ located around $x=0$. In this region, the convective flux is $J_c=u P_s\simeq \zeta^{-1} \mu x^2 P_s$ whereas the diffusive flux is of the order of $J_d=D_0\nabla P_s\simeq D_0 P_s/l^*$. The fluxes $J_c$ and $J_d$ are of the same order of magnitude when $x\sim l^*$ if we set the characteristic size $l^*$ to 
\begin{align}
	l^*=(k_BT/\mu)^{1/3}.
\end{align} 
The characteristic time  in the inner region is 
\begin{align}
	\tau^*=\frac{(l^*)^2}{D_0}=\frac{\zeta}{\mu^{2/3}(k_BT)^{1/3}}.
\end{align}
The time $\tau^*$ diverges in the small temperature limit, and we therefore refer to the region around $x=0$ as a {\em slow} region, as opposed to the outer region called the {\em fast} region. 

\begin{figure}[h!]
\includegraphics[width=7cm]{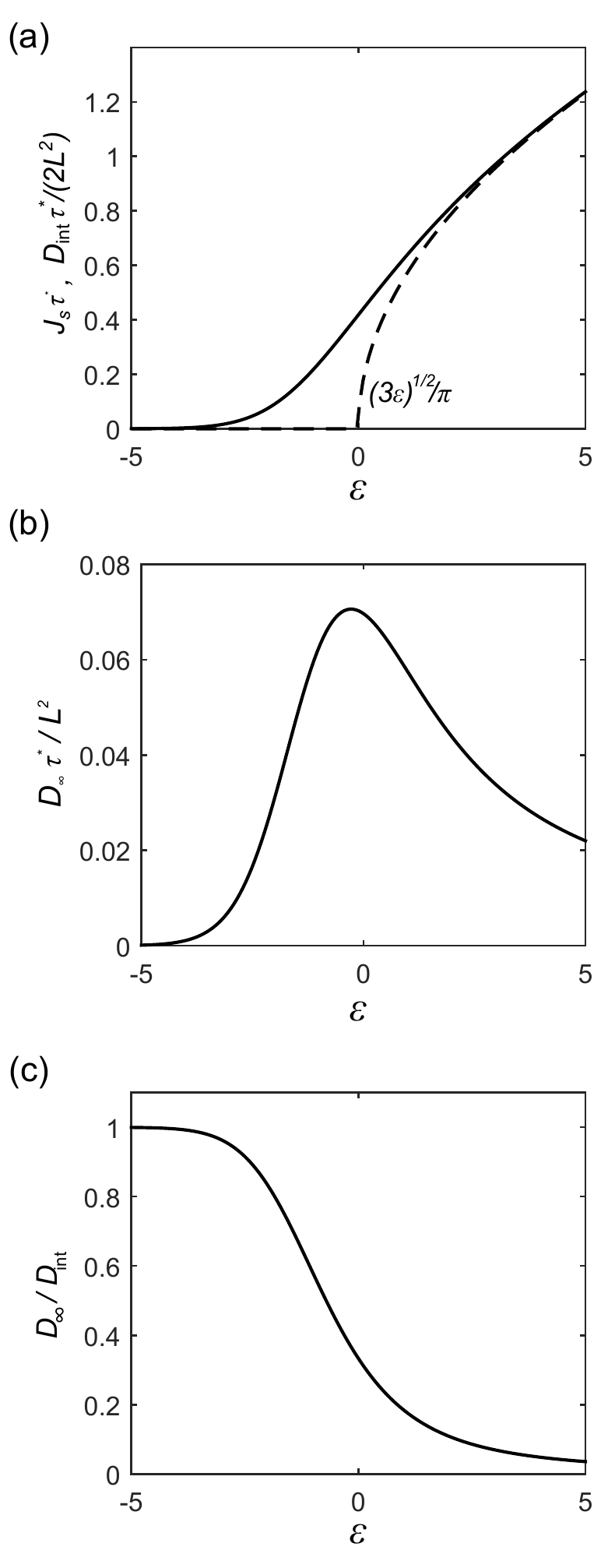} 
 \caption{(a) Value of the dimensionless flux $\tilde J_s=J_s \tau^*$ for near critical forces, which is also proportional to the effective diffusion coefficient in the intermediate time regime [see Eq.~(\ref{Dint})]. (b) Rescaled late time effective diffusivity $D_{\infty}$, deduced from Eq.~(\ref{Dinf}). (c) Ratio of the  late time effective diffusivity over the intermediate time effective diffusivity.    
\label{Fig2DiffCoeff}}
\end{figure}

Let us briefly derive the stationary probability density and steady state flux; although these quantities are known in the literature, we show below that they play an important role in the dispersion properties. Since each particle spends most of its time in one of the slow regions, the steady state stationary density $P_s(x)$ of positions (modulo $L$) is localized in these regions, where it satisfies
\begin{align}
&	-\zeta^{-1} \partial_x [(3 \mu x^2 +F-F_c) P_{s}]+ D_0\ \partial_x^2 P_{s}=0 \label{942},
\end{align}
where we have used a Taylor expansion of the potential $V(x)$ at next to leading order around $x=0$. We introduce the dimensionless position $y=x/l^*$, probability density $\tilde P_{s}(y)=P_{{s}}l^*$ and flux $\tilde J_{s}= \tau^* J_{s}$. 
Using these notations, integrating Eq. (\ref{942}) and comparing with Eq. (\ref{DefJs})  leads to
\begin{align}
	- [(3  y^2 +\varepsilon)  \tilde P_{s}]+ \partial_y \tilde P_{s}=-\tilde J_{s}.
\end{align}
Assuming that $\tilde P_{s}(y)$ vanishes for $y\rightarrow\infty$ (that is, for $x\gg l^*$), the solution of this equation is
\begin{align}
	\tilde P_{s}(y)=\tilde J_{{s}} \int_y^\infty du \ e^{-u^3-\varepsilon u+y^3+\varepsilon y}\label{ExpressionPstat}
\end{align}
The normalization imposes that $1=\int_{-L/2}^{L/2}dx P_{{s}}(x)\simeq \int_{-\infty}^{\infty}dy \tilde P_{{s}}(y)$ and thus leads to the identification of the dimensionless current: 
\begin{align}
	\tilde{J}_{s}^{-1}=\int_{-\infty}^{\infty} dy\int_0^{\infty}du  \   e^{-(u+y)^3-\varepsilon u+y^3} \label{FluxRescaled}.
\end{align}
The rescaled stationary PDF in the slow region is represented in Fig.\ref{Fig1}(b), where one observes the transition from narrow distributions shifted in the region $x<0$ when $\varepsilon<0$ (that is, for forces $F<F_c$) to broader and more centered distributions when $F>F_c$. The dimensionless flux is represented on Fig.\ref{Fig2DiffCoeff}(a), it almost vanishes for $F<F_c$ and then significantly increases when $F>F_c$. 

In the outer region, $\vert x\vert \gg l^*$, the convective term takes over the diffusive term in Eq.~(\ref{DefJs}), and we simply obtain 
\begin{align}
	P_s(x)\simeq J_s/u(x), \label{PsOuterRegion}
\end{align}
physically this means that in the regions where drift dominates, the probability of presence is inversely proportional to the speed.
We note that the expressions in the outer and the inner region [Eqs. (\ref{ExpressionPstat},\ref{PsOuterRegion})] can be matched in the region $l^*\ll \vert x\vert \ll L$, where both approximations give $P_s(x)\simeq J_s\zeta/(3 \mu x^2+F-F_c)$. 

\subsection{MSD at intermediate time scales:  ballistic and diffusive regime}
We now investigate the properties of the MSD $\psi(t)$ at different time scales. First, at very small times, $t\rightarrow0$ (or $s\rightarrow\infty$), we have  $\psi(t)=2D_0t$ where the effective diffusivity is exactly equal to the molecular diffusivity. We do not discuss this regime any further and consider now larger times, with the additional condition that $t \ll \tau^*$ (or equivalently $s \tau^*\gg1$). At this time scale, one can consider that the events of crossing the slow region, or escaping from it, take an infinite time.  

In this regime, when $x,x_0$ are in the {\em fast} region with $0<x_0<x<L$, we can approximate the FPT to $x$ from $x_0$  by the duration of the deterministic trajectory that links $x_0$ and $x$:
\begin{align}
	f(x,t\vert x_0) \simeq \delta\left(t-\int_{x_0}^x\frac{dy}{u(y)}\right).
\end{align}
In the opposite direction where $x_0>x$, reaching $x$ from $x_0$ requires crossing the slow region, which takes a time infinite compared to the time scale considered here. Hence $f(x,t\vert x_0)$ is simply approximated by $0$ in this case. Taking all this into account, the FPT density in Laplace space reads
\begin{align}
	\hat{f}(x,s\vert x_0)\simeq 
\begin{cases}
e^{-s\int_{x_0}^x\frac{dy}{u(y)}}& \ \mathrm{if } \ (0<x_0<x<L)\\
0 & \ \mathrm{if } \ (0<x<x_0<L)
\end{cases}\label{EqFPT}
\end{align}
We now calculate $\hat{f}_{s}(s,x)$ the average FPT density to $x$ with stationary initial conditions: 
\begin{align}
	\hat{f}_{s}(s,x)\simeq \int_{x_-}^{x} d{x_0} P_{s}(x_0)\hat{f}(s,x\vert x_0). 
\end{align}
Here the upper integration limit has been set to $x$, because if 
$x_0>x$ the FPT from $x_0$ to $x$ is infinite  at this time scale. In turn, the lower integration limit has been set to $x_-$, which is a position such that $l^*\ll x_-\ll L$ (its precise value will not change the result, see below). The reason for this is that if $x_0\sim l^*$ is in the slow region, its escape time is $\sim \tau^*$ and therefore almost no trajectory can bring it to $x$ at the considered time scale. 
For $x\gg l^*$, we can approximate $P_s(x_0)\simeq J_s/u(x_0)$ [see Eq.~(\ref{PsOuterRegion})] and therefore
\begin{align}
	\hat{f}_{s}(s,x)\simeq \int_{x_-}^{x} d{x_0} \frac{J_s}{u(x_0)} e^{-s\int_{x_0}^x\frac{dy}{u(y)}}.
\end{align}
The integration over $x_0$ is now straightforward:
\begin{align}
	\hat{f}_{s}(s,x)\simeq  \frac{J_s}{s}, \label{EqFPTs}
\end{align}
and the result does not depend on the lower bound of the integral $x_-$, as soon as $x_-\ll L$. 

\begin{figure}[h!]
\includegraphics[width=7cm]{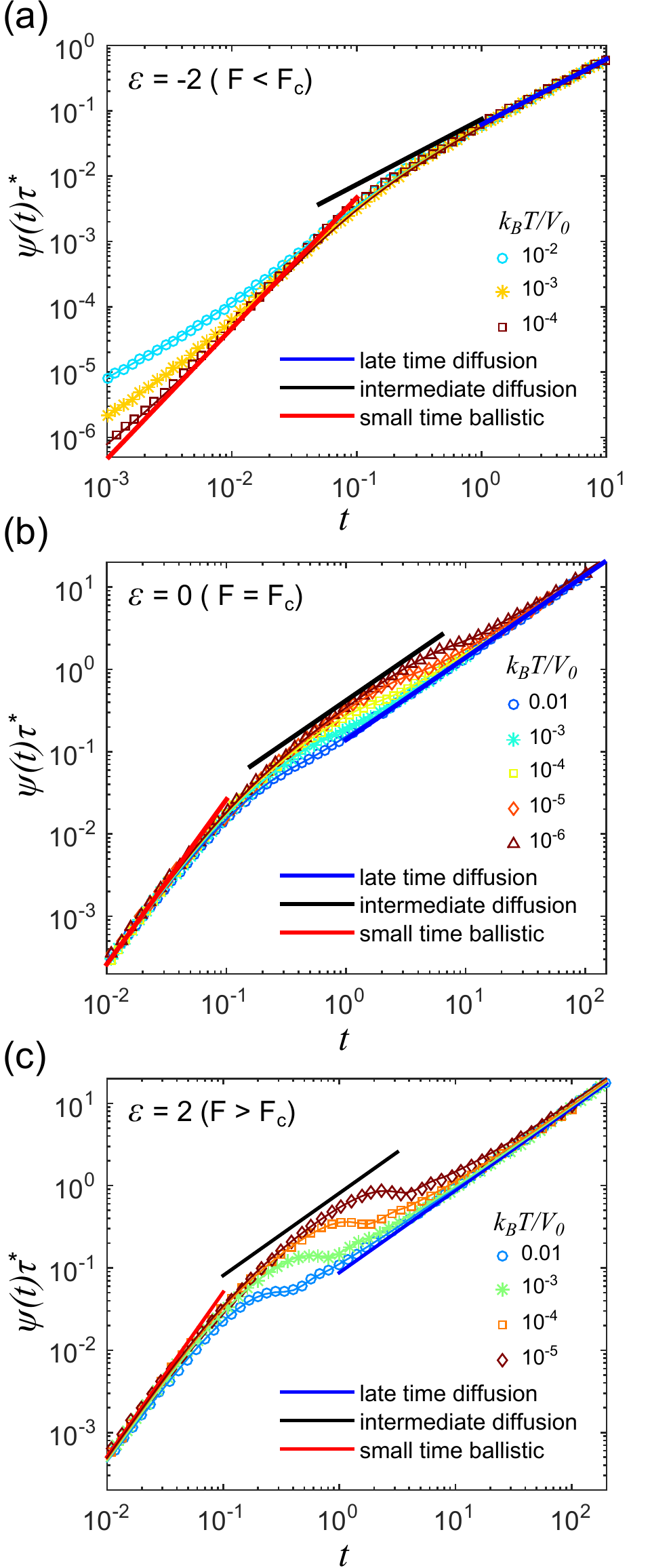} 
 \caption{MSD curves for the diffusion in a sine potential $V(x)=V_0\sin(2\pi x/L)$ for a tilting force  (a) below the critical force ($\varepsilon=-2$), (b) at the critical force ($\varepsilon=0$) and (c) above the critical force ($\varepsilon=2$). Different values of the noise amplitude are indicated in the legend. Symbols: results of the simulations of the Langevin equation (\ref{EqLangevin})  (see Appendix \ref{AppendixSimus} for details). Lines: results of the numerical evaluation of the Kubo formula (\ref{StartingEqMSD}). Bold lines: asymptotic results of Eq.~(\ref{SummaryResults}) (with no free parameters). Note that the regime $\psi=2D_0t$ is not represented but is present at short times. Units of length and times are $L$ and $\zeta L^2/V_0$, respectively. 
\label{FigSimus}}
\end{figure}  

Furthermore, when $x,x_0$ are in the ``fast'' regions, the approximation $u(x)P_s(x)=J_s=-u^*(x_0)P_s(x_0)$ holds [cf. Eq. (\ref{PsOuterRegion})]. Hence, inserting (\ref{EqFPT}) and (\ref{EqFPTs}) into Eq.~(\ref{StartingMSD}), and keeping only the dominant contribution to the MSD (coming from the $x,x_0$ in the fast region),  the expression of $\psi$ near criticality  is considerably simplified:
\begin{align}
\hat{\psi}(s)\simeq \frac{2J_{{s}}^2}{s^3}\int_0^{L}dx\int_0^{x}dx_0 \left[\frac{s\ e^{-s\int_{x_0}^x\frac{dy}{u(y)}}}{J_{{s}}}-1\right]. 
\end{align}
Here we investigate the regime $t\ll \tau^*$, or equivalently $s \tau^* \gg1$. Since $J_s\sim 1/\tau^*$, we obtain $s/J_s\gg1$  and
\begin{align}
\hat{\psi}(s)\simeq \frac{2J_{{s}}}{s^2}\int_0^{L}dx\int_0^{x}dx_0 \ e^{-s\int_{x_0}^x\frac{dy}{u(y)}}. \label{0403}
\end{align}
The above equation describes the MSD at time scales smaller than $\tau^*$. Let us consider its asymptotics. First, for $s\rightarrow\infty$, the terms $x\simeq x_0$ essentially contribute in the integral, and therefore
\begin{align}
\hat{\psi}(s)\simeq \frac{2J_{{s}}}{s^2}\int_0^{L}dx\int_0^{x}dx_0\  e^{\frac{-s(x-x_0)}{u(x)}}. 
\end{align}
Performing the integral, we obtain
\begin{align}
	\hat{\psi}(s)\simeq\frac{2J_{s}}{s^3}\int_0^{L}dx \ u(x)=\frac{2J_{s}F_c L}{\zeta s^3}. 
\end{align}
If we invert the Laplace transform, we obtain
\begin{align}
	\psi(t)= \frac{J_{s}F_c L} {\zeta} t^2,\label{ballisticMSD}
\end{align}
and thus the regime identified here is a ballistic regime, with a MSD $\psi(t) \simeq \alpha t^2$.  The non-trivial coefficient $\alpha$ identified in Eq.~(\ref{ballisticMSD}) is the product of two velocities, the first velocity is the average velocity $J_sL$, while the second  velocity is the characteristic velocity in the fast region $F_c/\zeta$. Note the non-trivial temperature dependence (as $(k_BT)^{1/3}$) of the coefficient of the MSD in this ballistic regime. 


Now, let us take the small $s$ limit of Eq.~(\ref{0403}), for which
\begin{align}
 \hat{\psi}(s)=\frac{2J_{{s}}}{s^2}\int_0^{L}dx \ x \simeq \frac{L^2 J_s}{s^2}.
\end{align}
Inverting the Laplace transform leads to $\psi(t)=2 D_\mathrm{int} t$ with $D_\mathrm{int}$ an effective diffusion coefficient at intermediate times, 
\begin{align}
D_\mathrm{int} =L^2 J_s  /2 \label{Dint}
\end{align}
Thus, the analysis reveals the existence of a diffusive regime at intermediate time scales. The effective diffusivity  $D_\mathrm{int}$ is proportional to the flux of particles and therefore  increases significantly when the force becomes larger than the critical force [that is, for increasing $\varepsilon$, see Fig.~\ref{Fig2DiffCoeff}(a)]. 

When $t\gg \tau^*$, the motion becomes effectively diffusive with an effective diffusion coefficient $D_{\infty}=(L^2/\tau^*) G(\varepsilon)$, where $G$ is the same dimensionless function identified in Refs.~\cite{reimann2002diffusion,reimann2001giant}. It is represented in Fig. \ref{Fig2DiffCoeff}(b), and shows a maximum for $\varepsilon\simeq0$, leading to the giant enhancement of diffusivity at the critical force. For completeness we provide in Appendix \ref{AppendixDinfini} a derivation of $D_{\infty}$ within our formalism, which is an alternative to the approach of Refs.~ \cite{reimann2002diffusion,reimann2001giant}. 

The ratio of the late time over the intermediate time diffusivity 
 $D_\infty/D_{\mathrm{int}}$ is shown on Fig. \ref{Fig2DiffCoeff}(c). It is almost equal to $1$ for forces below the critical force ($\varepsilon<0$) but then vanishes for larger values of $\varepsilon$. This means that, when the critical force is reached, the shape of the MSD curves changes drastically. For $F<F_c$, one observes a direct transition between a ballistic and the long time ballistic regime. When $F>F_c$, one observes the intermediate diffusive regime with a diffusivity larger than the effective diffusivity, which translates by an  {\em overshoot} of the MSD and an apparent regime of subdiffusion.

If we summarize all the results, we obtain 
\begin{align}
\psi(t)\simeq
\begin{cases}
2D_0 t & \text{if} \ t\ll t_2\\
J_{s}v L t^2 & \text{if} \ t_2\ll t\ll L/v\\
2D_{\mathrm{int}} t & \text{if} \ L/v\ll t\ll \tau^* \\
2D_{\infty} t & \text{if} \ \tau^*\ll t
\end{cases}\label{SummaryResults}
\end{align}
with $v=F_c/\zeta$ is the velocity in the fast region and $t_2=k_BT/(F_c L J_s)$. In order to check these predictions, we performed stochastic simulations of the Langevin equation (\ref{EqLangevin}) in the case of a sine potential (see Appendix \ref{AppendixSimus} for details on the simulation algorithm). The results presented in Fig.\ref{FigSimus} confirm the validity of the asymptotic regimes described by Eq.~(\ref{SummaryResults}) for tilting forces that are either below, above or equal to the critical force.


\section{Conclusion}

In this paper we have studied the time dependent dispersion properties of particles diffusing in a near critically tilted one-dimensional periodic potential. We have derived explicit asymptotic expressions for the MSD at different time scales [Eq.~(\ref{SummaryResults})]. The approach to the late time diffusive limit depends only on a small number of parameters which characterize the potential in particular regions where the dynamics is slow. The approach to the diffusive limit is therefore universal in the sense that it does not depend on the details of the potential shape. 

It has been proposed that the giant increase of the late time diffusivity that occurs at the critical force \cite{reimann2001giant} can be used to estimate barriers of potential energy \cite{hayashi2015giant,Evstigneev2008}. Here we have shown that the time dependent dispersion properties are strongly modified when the force becomes larger than $F_c$, in particular we have found the presence of a second diffusive regime at intermediate time scales. The strong difference between these two diffusion coefficients could be used as another signature of the effect of the crossing of the critical force.  

In this study, we have quantified the transition between the short time regime of molecular diffusion and the late time effective diffusion. The Mean-Square-Displacement  between these regimes can be regarded as ``anomalous'' , in the sense that it is a non-linear function of time. Anomalous diffusion can have a variety of origins in different physical systems and may occur in the late time regime for example for particles undergoing jumps whose sizes or durations follow large distributions \cite{Metzler2000}, when the distribution of energy barriers leads to  diverging mean occupation times in local energy minima, or when the convective velocity field has long range correlations \cite{BOUCHAUD1990}. Anomalous diffusion also arises in fractal media \cite{benAvraham2000},  when considering time dependent microscopic diffusivities \cite{jeon2014scaled},  or when the tracer trajectory results from a collective dynamics, such as in polymer systems \cite{DoiEdwardsBook,Panja2010} or complex fluids \cite{jeon2013anomalous,wei2000single}. 
This work, where we study out-of-equilibrium tracer particles in periodic media, is an example where one can entirely characterize the ``anomalous''  time-dependent dispersion properties at all intermediate time scales between the molecular diffusion regime and the final late time regime, which is  however one of normal diffusion.

\appendix
\section{Late time effective diffusivity}
\label{AppendixDinfini}

In this appendix we briefly derive an expression for the late time effective diffusivity $D_\infty$. If we expand the temporal Laplace transform of the FPT densities for $s\rightarrow0$ we obtain
\begin{align}
	\hat{f}(x,s\vert x_0)\simeq1-\tau(x\vert x_0) s + ...\\
	\hat{f}_{{s}}(x,s)\simeq1-\tau_{{s}}(x) s + ... 
\end{align}
where $\tau(x\vert x_0)=\int_0^\infty dt f(x,t\vert x_0)t $ is the Mean First Passage Time (MFPT) to reach the position $x$ modulo $L$ starting from $x_0$, while $\tau_s(x)$ is the MFPT to $x$ with $x_0$ averaged over stationary initial conditions. Using Eq.~(\ref{StartingMSD}), we then find that $\hat\psi\simeq2 D_\infty/s^2$, where the effective diffusivity  $D_{\infty}$ reads
\begin{align}
D_{\infty}=D_0+\int_0^{L}dx\int_0^{L}dx_0 \ u(x)P_{{s}}(x)   \nonumber\\
 u^*(x_0)P_{s}(x_0)\left[\tau(x\vert x_0)-\tau_{{s}}(x)\right]&.\label{GenExpressionDinfty}
\end{align}
For forces close to the critical force, this  expression can be considerably simplified. 
The dominant contribution to this integral comes from the values of $x,x_0$ that are outside the slow region, in which the diffusive component of the flux is negligible, and thus $u(x)P_s(x)=J_s=-u^*(x_0)P_s(x_0)$ [cf. Eq. (\ref{PsOuterRegion})]. Furthermore, taking $x$ and $x_0$ in $[0,L]$, we realize that $\tau(x\vert x_0)$ is negligible when $x_0<x$ (because the convection brings the particle to $x$ almost immediately) whereas for $x<x_0$ this time is equal to the time to cross the slow region, which is exactly the inverse of the flux $J_s$. 

We define $\tau_e$ the time to escape the slow region starting from stationary initial conditions, and we note that $\tau_{s}(x)\simeq\tau_e $ is independent of $x$.
Following these considerations,  and neglecting the term $D_0$ in Eq.~(\ref{GenExpressionDinfty}), we obtain
\begin{align}
D_{\infty}\simeq J_s^2 L^2 \left[\frac{1}{2J_s}-\tau_e\right].
\end{align}
Now let us consider $\tau_{\infty}(x_0)$, the mean time to escape the slow region, starting from $x_0$, from which $\tau_e=\langle  \tau_{\infty}(x_0)\rangle$ can be deduced. 
We pose $y_0=x_0/l^*$ and $\tau_{\infty}(x_0)=\tau^* \tilde\tau_{\infty}$. Then, the dimensionless MFPT satisfies the backward equation \cite{VanKampen1992,gardiner1983handbook}
\begin{align}
	(3 y_0^2+\varepsilon)\partial_{y_0}\tilde \tau_{\infty}+\partial_{y_0}^2 \tilde\tau_{\infty}=-1 \label{ODEtauInfty}
\end{align}
Noting that
\begin{align}
	\partial_{y_0}[e^{y_0^3+\varepsilon y_0}\partial_{y_0}\tilde\tau_{\infty}]=-e^{y_0^3+\varepsilon y_0}
\end{align}
the differential equation (\ref{ODEtauInfty}) can be integrated twice, leading to 
\begin{align}
	\tilde\tau_{\infty}(y_0)=\int_{y_0}^{\infty} dz \int_0^{\infty}du \ e^{-z^3+(z-u)^3-\varepsilon u}\label{7574}
\end{align}
where we took into account the condition that $\tilde\tau_{\infty}(y_0)$ vanishes at infinity.  Now, the average escape time 
from the slow region can be calculated by averaging over $P_s$ given in Eq.~(\ref{ExpressionPstat}), leading to
\begin{align}
	\tilde \tau_e&=\tilde{J}_{s} \int_{-\infty}^{\infty}dy \int_0^{\infty} dv \int_0^{\infty} dw \int_0^{\infty} du \nonumber\\
 &e^{-(v+y)^3-\varepsilon v+y^3}   e^{-(y+w)^3+(y+w-u)^3-\varepsilon u}\label{taue}
\end{align}
Taking $y_0\rightarrow-\infty$ in  (\ref{7574}) we can check that the average time to cross the slow region is also $1/J_s$. Therefore, we obtain at the end
\begin{align}
D_{\infty}\simeq \frac{L^2}{\tau^*} \tilde J_s^2 \left[\frac{1}{2\tilde J_s}-\tilde \tau_e \right]=\frac{L^2}{\tau^*}G(\varepsilon), \label{Dinf}
\end{align}
where the dimensionless function $G(\varepsilon)$ is deduced from Eqs.(\ref{FluxRescaled},\ref{taue}) and is represented on Fig.~\ref{Fig2DiffCoeff}, and is in excellent agreement with the results of Refs.~\cite{reimann2002diffusion,reimann2001giant}.

\section{Details on simulations}
\label{AppendixSimus}
We performed numerical simulations of the Langevin equation 
(\ref{EqLangevin}) by using the algorithm $x_{i+1}=x_i+u(x_i)\Delta+\sqrt{2\Delta D_0}w_i$, with $w_i$ a Gaussian random variable of zero mean and variance $1$, $\Delta$ the time step and $x_i$ the position at time $t=i\Delta$. Each trajectory was simulated during a time $t_m$, the position of the particle at the end of a trajectory being used as the initial position for the next trajectory. The MSD was then computed with Eq. (\ref{DefMSD}) by averaging over distinct trajectories. For each parameter, $\Delta$ was chosen small enough so that we could observe the regime $\psi(t)\simeq2D_0t$ for small times (not shown in Fig.~\ref{FigSimus}), and we carefully controlled that MSD curves obtained with different $\Delta$ overlapped. 
In order to ensure that different runs are independent, the time $t_m$ was chosen large enough to be located in the late time diffusive regime. The number of runs was always larger $8,000$ for each parameter set. In the late time regime, where $X(t)-X(0)$ is Gaussian distributed, the $95\%$ confidence intervals  can be evaluated to $(1\pm\sqrt{8/N_{\mathrm{run}}})\psi$ so that the precision on $\psi$ is approximately $\pm3\%$.


\end{document}